# Causality and phase transitions in $\mathcal{PT}$-symmetrical optical systems

A. A. Zyablovsky,[1,2,3] A. P. Vinogradov,[1,2,3] A. V. Dorofeenko,[1,2,3] A. A. Pukhov,[1,2,3] and A. A. Lisyansky[4]

[1]All-Russia Research Institute of Automatics, 22 Sushchevskaya, Moscow 127055, Russia
[2]Moscow Institute of Physics and Technology, Moscow Region, Dolgoprudny, Russia
[3]Institute for Theoretical and Applied Electromagnetics RAS, Moscow, Russia
[4]Department of Physics, Queens College of the City University of New York, Queens, NY 11367, USA

We discuss phase transitions in $\mathcal{PT}$-symmetric optical systems. We show that due to frequency dispersion of the dielectric permittivity, an optical system can have $\mathcal{PT}$-symmetry at isolated frequency points only. An assumption of the existence of a $\mathcal{PT}$-symmetric system in a continuous frequency interval violates the causality principle. Therefore, the ideal symmetry-breaking transition cannot be observed by simply varying the frequency.

In the last decade, there has been rising interest in optics of artificial materials. Among such materials, the systems with balanced loss and gain regions have attracted particular attention [1-3]. The concept of these systems stems from the idea of the extension of quantum mechanics to non-Hermitian Hamiltonians possessing $\mathcal{PT}$-symmetry [4, 5]. $\mathcal{PT}$-symmetrical systems are invariant with respect to the simultaneous spatial inversion and time inversion. The former performed by the linear operator $\hat{\mathcal{P}}$ which transforms coordinates and momenta as $\mathbf{r} \to -\mathbf{r}$ and $\mathbf{p} \to -\mathbf{p}$, while the time inversion is performed by the antilinear operator $\hat{\mathcal{T}}$, which transforms $\mathbf{p} \to -\mathbf{p}$ and $i \to -i$; the simultaneous application of these operators, $\hat{\mathcal{P}}\hat{\mathcal{T}}$, is antilinear, it transforms $\mathbf{r} \to -\mathbf{r}$, $\mathbf{p} \to \mathbf{p}$ and $i \to -i$.

Among the intriguing properties of $\mathcal{PT}$-symmetric optical structures are double refraction [2] and nonreciprocal diffraction patterns [2, 6, 7], power oscillations [2, 6], loss induced optical transparency [8], coherent perfect absorber - laser [9, 10], nonlinear switching systems [11], nonreciprocal Bloch oscillations [12, 13], unidirectional invisibility [13-16] and breaking of $\mathcal{PT}$-symmetry of eigensolutions without breaking the $\mathcal{PT}$-symmetry of the system [1-3, 6]. The latter phenomenon is called a phase transition and has been predicted for one- [10] and two-dimensional [1, 2] systems. However, an analysis of the conditions for observation of $\mathcal{PT}$-symmetry breaking transitions in real systems has not been carried out.

In optics, $\mathcal{PT}$-symmetry is usually studied in the frequency domain by considering solutions of the scalar Helmholtz equation for the z-component of the electric field $E$:

$$\left(\frac{\partial^2}{\partial x^2} + \frac{\partial^2}{\partial y^2} + \frac{\omega^2}{c^2}\varepsilon(\omega,x,y)\right)E(\omega,x,y) = 0, \qquad (1)$$

where $\varepsilon(\omega, x, y)$ is dielectric permittivity of a non-magnetic medium. For one-dimensional systems, both $\varepsilon$ and $E$ should depend only on one spatial coordinate. In this case, the $\mathcal{PT}$-transformation reduces to the reflection in a plane crossing the $z$-axis and complex conjugation.

A system is $\mathcal{PT}$-symmetrical if and only if Eq. (1) is invariant with respect to the $\hat{\mathcal{P}}\hat{\mathcal{T}}$-transformation. This happens when $\varepsilon(\omega, x)$ satisfies the condition [1]

$$\operatorname{Re}\varepsilon(\omega, x) = \operatorname{Re}\varepsilon(\omega, -x), \tag{2}$$

$$\operatorname{Im}\varepsilon(\omega, x) = -\operatorname{Im}\varepsilon(\omega, -x). \tag{3}$$

Thus, such a system includes both absorbing and amplifying media (we exclude the trivial case of $\operatorname{Im}\varepsilon(\omega, x) = 0$).

Due to anti-linearity of the $\hat{\mathcal{P}}\hat{\mathcal{T}}$-operator, the eigensolutions of Eq. (1) may or may not be $\mathcal{PT}$-symmetric depending on the values of permittivity [6, 17]. When parameters of the systems are varied, one or multiple transitions between phases with $\mathcal{PT}$-symmetric and non-symmetric eigensolutions may happen [1-3, 6]. In any $\mathcal{PT}$-symmetric system, the $\mathcal{PT}$-symmetric eigensolutions have real eigenvalues and the eigensolutions with broken $\mathcal{PT}$-symmetry have complex eigenvalues [1-3]. Therefore, to study the symmetry breaking phase transition, it is sufficient to trace the behavior of eigenvalues. In one-dimension, these eigenvalues are the eigenvalues of the scattering matrix [10]; in two-dimensions, one can consider the wavenumbers of eigensolutions of Eq. (1) [1-3].

In realistic optical systems, $\mathcal{PT}$-symmetry breaking transitions are not easy to achieve, because to satisfy Eqs. (2) and (3) one must simultaneously tune permittivities for both absorbing and amplifying media. If pumping rate is intended as a tuning parameter, both media should have population inversions so that they are affected by pumping. As a consequence, the pumping should depend on space coordinates. For example, it should be greater in gain regions and smaller in absorbing regions. It seems that the easiest way for observing these phase transitions is by varying the frequency of the external electric field. As we show below, this option is not available because an assumption of the existence of a $\mathcal{PT}$-symmetric system in a continuous frequency interval violates the causality principle.

In this Communication we prove that due to dispersion of the dielectric function, the $\mathcal{PT}$-symmetry may exist in optical systems only at isolated frequencies. Therefore, the ideal symmetry-breaking transition cannot be observed by simply varying the frequency.

In any optical system with either loss or gain, frequency dispersion of the permittivity is crucial. Due to causality, a dielectric function $\varepsilon(\omega, x)$ must be analytic in the upper half of the complex frequency plane so that all its singularities are situated in the lower half of the complex plane [18]. Causality must hold for both dissipative and active systems.

Due to the causality principle, $\varepsilon(\omega,x)$ must satisfies the Kramers-Kronig relations

$$\operatorname{Re}\varepsilon(\omega,x) = \varepsilon_0 + \frac{1}{\pi} v.p. \int_{-\infty}^{+\infty} \frac{\operatorname{Im}\varepsilon(\omega',x)}{\omega'-\omega} d\omega', \qquad (4)$$

$$\operatorname{Im}\varepsilon(\omega,x) = -\frac{1}{\pi} v.p. \int_{-\infty}^{+\infty} \frac{\operatorname{Re}\varepsilon(\omega',x)-\varepsilon_0}{\omega'-\omega} d\omega', \qquad (5)$$

where $\varepsilon_0$ is the permittivity of vacuum. Using these relations one can show that conditions (2) and (3) can be satisfied for a discrete set of frequencies only. Indeed, if the $\mathcal{PT}$-symmetry condition (3) holds for any real values of $\omega$, then Eq. (4) provides that

$$\operatorname{Re}\varepsilon(\omega,-x) = \varepsilon_0 + \frac{1}{\pi} v.p. \int_{-\infty}^{+\infty} \frac{\operatorname{Im}\varepsilon(\omega',-x)}{\omega'-\omega} d\omega' = \varepsilon_0 - \frac{1}{\pi} v.p. \int_{-\infty}^{+\infty} \frac{\operatorname{Im}\varepsilon(\omega',x)}{\omega'-\omega} d\omega'. \qquad (6)$$

Now, as it follows from Eqs. (4) and (6), in order to satisfy Eq. (1), it is necessary that

$$v.p. \int_{-\infty}^{+\infty} \frac{\operatorname{Im}\varepsilon(\omega',x)}{\omega'-\omega} d\omega' = 0. \qquad (7)$$

This is only possible for a completely transparent system with $\operatorname{Re}\varepsilon(\omega,x) = \varepsilon_0$ and $\operatorname{Im}\varepsilon(\omega,x) = 0$. Thus, a physical system cannot possess properties (2)-(3) for an infinite frequency interval.

The impossibility of existence of a $\mathcal{PT}$-symmetry in a frequency range has a simple mathematical reason. Let us denote the dielectric functions in two points $x$ and $-x$ as $\varepsilon_1(\omega)$ and $\varepsilon_2(\omega)$, respectively. Let us presume that $\varepsilon_2(\omega)$ is a "causal" dielectric function, so it is analytic in the upper half of the complex plane. For real frequencies, Eqs. (2) and (3) are equivalent to complex conjugation. Since complex conjugation does not preserve analyticity, $\varepsilon_2^*(\omega)$ cannot be used for $\varepsilon_1(\omega)$. The analytical function that satisfies Eqs. (2) and (3) is

$$\varepsilon_1(\omega) = \varepsilon_2^*(\omega^*). \qquad (8)$$

However, all singularities of $\varepsilon_1(\omega)$ defined by Eq. (8) are in the upper half-plane, and therefore, the respective response function would violate causality

Note, that $\mathcal{PT}$-symmetry for *all* frequencies is not required for observing the $\mathcal{PT}$-symmetry transition with varying frequency. It would suffice for Eqs. (2)-(3) to be valid in a finite frequency range. However, this also is not possible. As it follows from the identity theorem [7], two analytical functions coincide in the upper half-plane if they coincide on any open interval of the real axis. The identity (8) must be true in the case of finite interval as well. Thus, a dielectric function, which is $\mathcal{PT}$-symmetrical over a finite frequency interval would not satisfy the Kramers-Kronig relations and would violate causality.

As an example, let us consider a medium described by dielectric permittivity of with the Lorentzian-shape line centered at the frequency $\omega_0$:

$$\varepsilon(\omega) = \varepsilon_m - \frac{\alpha}{\omega^2 - \omega_0^2 + 2i\gamma\omega}, \qquad (9)$$

where $\varepsilon_m$ is a dielectric permittivity of the matrix in which amplifying (absorbing) media are placed, $\alpha$ and $\gamma$ are the strength and the linewidth of amplification (absorption), respectively. For an absorbing medium $\mathrm{Im}\,\varepsilon > 0$, thus $\alpha$ and $\gamma$ must be positive. For an amplifying medium, $\mathrm{Im}\,\varepsilon < 0$, so that one of the parameters $\alpha$ or $\gamma$ must be negative. For real frequencies, negative $\gamma$ is equivalent to the complex conjugation of $\varepsilon(\omega)$. This corresponds to moving the pole of $\varepsilon(\omega)$ from the lower to the upper half of the complex frequency plane. The latter violates causality for the response function. Therefore, the only choice for parameters $\alpha$ and $\gamma$ is $\alpha < 0$ and $\gamma > 0$, which corresponds to the "antiresonance" of the real part of the dielectric function [20]. The choice $\alpha < 0$ and $\gamma > 0$ ensures causality but is incompatible with $\mathcal{PT}$-symmetry. Indeed, as shown in Fig. 1, when condition (1) is satisfied, due to resonant behavior of the absorbing medium and antiresonant behavior of the amplifying medium, condition (2) can hold at one point, $\omega = \omega_0$, only. This illustrates the general rule following from the Kramers-Kronig relations.

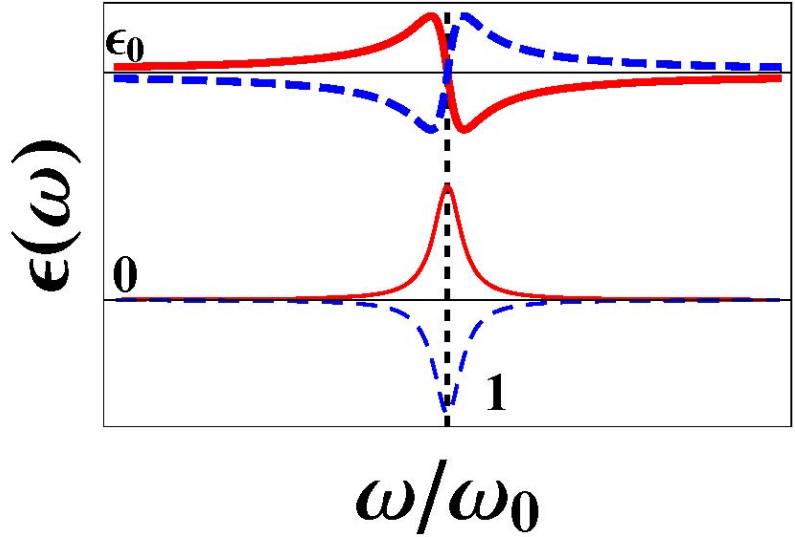

FIG 1. Real (thick lines) and imaginary (thin lines) parts of dielectric permittivity of absorbing (solid lines) and amplifying (dashed lines) media dielectric permittivities are given by Eq. (9) with opposite signs of the imaginary parts.

Another type of symmetry breaking phase transitions may occur in a non-$\mathcal{PT}$-symmetric system, which can be converted to the $\mathcal{PT}$-symmetric system by a formal, coordinates-dependent change of variables. An example of such a system [8] is an optical structure consisting of two parallel waveguides. The first waveguide is neither active nor lossy, $\operatorname{Im}\varepsilon_1 = 0$, the second one has losses, $\operatorname{Im}\varepsilon_2 = \alpha > 0$.

Assuming that waveguides are situated in the $xz$-plane and their axes are directed along the $x$-axis, one can introduce a "new" field $e(z,x) = E(z,x)\exp(\alpha z/2)$. The field $e(z,x)$ is governed by Eq. (1) with effective permittivity whose imaginary part $\operatorname{Im}\varepsilon^{eff}$ is equal to $-\alpha/2$ and $\alpha/2$ in the first and second waveguides, respectively.

Varying the parameter $\alpha$ one may observe a phase transition between the $\mathcal{PT}$-symmetric and non-$\mathcal{PT}$-symmetric distributions of the field $e(z,x)$ [8]. At the same time, since the field $e(z,x)$ is directly related to the real field $E(z,x)$, in the primary system, a "hidden" phase transition also occurs [8]. Note, that even if the transformed system has real eigenvalues, the eigenvalues of the original system are complex.

The effective permittivity may not be a response function, therefore it does not have to satisfy Kramers-Kronig relations. Thus, it may seem that limitations due to dispersion would not affect the possibility of satisfying conditions (2) and (3). We show, however, that dispersion of the permittivity of the real system does not allow for the "hidden" $\mathcal{PT}$-symmetry breaking phase transition in the described system.

In Ref. [8], the problem of determining the field distribution was considered in the coupling mode approximation and reduced to the system of equations for the optical mode [1]

$$i\frac{d}{dz}\begin{pmatrix}a_1\\a_2\end{pmatrix} = \begin{pmatrix}\beta_1+\delta & \kappa \\ \kappa^* & \beta_2+\delta^*\end{pmatrix}\begin{pmatrix}a_1\\a_2\end{pmatrix}, \qquad (10)$$

where $a_1$ and $a_2$ are amplitudes of the electric field in waveguides, $\beta_1 = (\omega/c)\sqrt{\varepsilon_1}$ and $\beta_2 = (\omega/c)\sqrt{\varepsilon_2}$ are wavenumbers in uncoupled waveguides, $\kappa$ is the complex coupling coefficient between waveguides, and $\delta$ is the change of the wavenumber due to the interaction between the waveguides. The amplitudes in the waveguides are the same if

$$(\beta_1 - \beta_2 + \delta - \delta^*) = (\kappa^* e^{-i\phi} - \kappa e^{i\phi}). \qquad (11)$$

where $\phi$ is a relative phase between fields in different waveguides. Since the right hand part of Eq. (11) is purely imaginary, the field distribution can be symmetrical only if $\beta_1 - \beta_2$ is also imaginary. Introducing $\Delta\beta = (\beta_1 - \beta_2)/2$, we can see that condition (11) can be satisfied if

$$\left|\operatorname{Im}(2\Delta\beta + \delta - \delta^*)\right| \leq 2|\kappa|, \qquad (12)$$

$$\operatorname{Re}\Delta\beta = 0. \qquad (13)$$

Thus, for eigensolutions to be symmetrical, the real part of permittivity must be an even function of the coordinate $\operatorname{Re}\varepsilon_1 = \operatorname{Re}\varepsilon_2$ or

$$\operatorname{Re}\varepsilon(\omega,x) = \operatorname{Re}\varepsilon(\omega,-x). \tag{14}$$

Generalizing Eq. (6) we obtain

$$\operatorname{Re}\varepsilon(\omega,x) = \varepsilon_0 + \frac{1}{\pi} v.p. \int_{-\infty}^{+\infty} \frac{\operatorname{Im}\varepsilon(\omega',x)}{\omega'-\omega} d\omega', \tag{15}$$

$$\operatorname{Re}\varepsilon(\omega,-x) = \varepsilon_0 + \frac{1}{\pi} v.p. \int_{-\infty}^{+\infty} \frac{\operatorname{Im}\varepsilon(\omega',-x)}{\omega'-\omega} d\omega'. \tag{16}$$

As it follows from Eqs. (15) and (16), in order to satisfy condition (14), one must have

$$v.p. \int_{-\infty}^{+\infty} \frac{\operatorname{Im}\varepsilon(\omega',x) - \operatorname{Im}\varepsilon(\omega',-x)}{\omega'-\omega} d\omega' = 0. \tag{17}$$

Equation (17) must hold for an arbitrary frequency (or at least, for a frequency from a continuous interval). This is only possible for the system with $\operatorname{Im}\varepsilon(\omega',x) - \operatorname{Im}\varepsilon(\omega',-x) = 0$. Then conditions (12) and (13) are always satisfied and all solutions are symmetrical. Following the proof for the exact $\mathcal{PT}$-symmetric systems presented above we obtain that if the value of $\operatorname{Im}\varepsilon(\omega',x) - \operatorname{Im}\varepsilon(\omega',-x)$ is not zero, then condition (14) can be satisfied for a discrete set of frequencies only.

Thus, due to frequency dispersion of permittivity in optical systems, ideal transitions (including "hidden" transitions) between $\mathcal{PT}$-symmetric and non-symmetric phases cannot be observed by simply varying the frequency. At the same time, there are no limitations for such transitions to occur by varying pump rate [1-3, 6, 8].

This work was partly supported by RFBR Grants Nos. 12-02-01093, 13-02-00407, 13-02-92660, by the Dynasty Foundation, and by the NSF under Grant No. DMR-1312707.

[1] R. El-Ganainy, K.G. Makris, D.N. Christodoulides, Z.H. Musslimani, Theory of coupled optical PT-symmetric structures, Opt. Lett., 32 (2007) 2632-2634.
[2] K.G. Makris, R. El-Ganainy, D.N. Christodoulides, Z.H. Musslimani, Beam Dynamics in PT Symmetric Optical Lattices, Phys. Rev. Lett., 100 (2008) 103904.
[3] S. Klaiman, U. Günther, N. Moiseyev, Visualization of Branch Points in PT-Symmetric Waveguides, Phys. Rev. Lett., 101 (2008) 080402.
[4] C.M. Bender, S. Boettcher, Real Spectra in Non-Hermitian Hamiltonians Having PT Symmetry, Phys. Rev. Lett., 80 (1998) 5243-5246.
[5] C.M. Bender, D.C. Brody, H.F. Jones, Complex Extension of Quantum Mechanics, Phys. Rev. Lett., 89 (2002) 270401.

[6] C.E. Rüter, K.G. Makris, R. El-Ganainy, D.N. Christodoulides, M. Segev, D. Kip, Observation of parity-time symmetry in optics, Nat. Phys., 6 (2010) 192-195.
[7] L. Feng, M. Ayache, J. Huang, Y.-L. Xu, M.-H. Lu, Y.-F. Chen, Y. Fainman, A. Scherer, Nonreciprocal Light Propagation in a Silicon Photonic Circuit, Science, 333 (2011) 729-733.
[8] A. Guo, G.J. Salamo, D. Duchesne, R. Morandotti, M. Volatier-Ravat, V. Aimez, G.A. Siviloglou, D.N. Christodoulides, Observation of PT-Symmetry Breaking in Complex Optical Potentials, Phys. Rev. Lett., 103 (2009) 093902.
[9] S. Longhi, PT-symmetric laser absorber, Phys. Rev. A, 82 (2010) 031801(R).
[10] Y.D. Chong, L. Ge, A.D. Stone, PT-Symmetry Breaking and Laser-Absorber Modes in Optical Scattering Systems, Phys. Rev. Lett., 106 (2011) 093902.
[11] A.A. Sukhorukov, Z. Xu, Y.S. Kivshar, Nonlinear suppression of time reversals in PT-symmetric optical couplers, Phys. Rev. A, 82 (2010) 043818.
[12] S. Longhi, Bloch Oscillations in Complex Crystals with PT Symmetry, Phys. Rev. Lett., 103 (2009) 123601.
[13] A. Regensburger, C. Bersch, M.-A. Miri, G. Onishchukov, D.N. Christodoulides, Parity-time synthetic photonic lattices, Nature, 488 (2012) 167-171.
[14] Z. Lin, H. Ramezani, T. Eichelkraut, T. Kottos, H. Cao, D.N. Christodoulides, Unidirectional Invisibility Induced by PT-Symmetric Periodic Structures, Phys. Rev. Lett., 106 (2011) 213901.
[15] X. Yin, X. Zhang, Unidirectional light propagation at exceptional points, Nat Mater, 12 (2013) 175-177.
[16] L. Feng, Y.-L. Xu, W.S. Fegadolli, M.-H. Lu, J.E.B. Oliveira, V.R. Almeida, Y.-F. Chen, A. Scherer, Experimental demonstration of a unidirectional reflectionless parity-time metamaterial at optical frequencies, Nat Mater, 12 (2013) 108-113.
[17] O. Bendix, R. Fleischmann, T. Kottos, B. Shapiro, Exponentially Fragile PT Symmetry in Lattices with Localized Eigenmodes, Phys. Rev. Lett., 103 (2009) 030402.
[18] L.D. Landau, E.M. Lifshitz, L.P. Pitaevskii, Electrodynamics of Continuous Media, Butterworth-Heinemann, Oxford, 1995.
[19] M.J. Ablowitz, A.S. Fokas, Complex variables: Introduction and applications, Cambridge University Press, Cambridge, 1997.
[20] S. Solimeno, B. Crosignani, P. DiPorto, Guiding, Diffraction, and Confinement of Optical Radiation, Academic Press, Orlando, 1986.